\documentclass[aps, showpacs, showkeys,nofootinbib,floatfix]{revtex4}

\usepackage{amssymb}
\usepackage{amsmath}
\usepackage{graphicx}

\begin{document}

\title{Geometrical Diagnostic for Generalized Chaplygin Gas Model}

\author{Jianbo Lu}
\email{lvjianbo819@163.com}
\author{Lixin Xu}
\email{lxxu@dlut.edu.cn}

\affiliation{School of Physics and Optoelectronic Technology, Dalian
University of Technology, Dalian, 116024, P. R. China}

\begin{abstract}

A new diagnostic method, $Om$   is applied to generalized Chaplygin
gas (GCG) model as the unification of dark matter and dark energy.
On the basis of the recently observed data: the Union supernovae,
the observational Hubble data, the SDSS baryon acoustic peak and the
five-year WMAP shift parameter, we show the discriminations between
GCG and $\Lambda$CDM model. Furthermore, it is calculated that the
current equation of state  of dark energy  $w_{0de}=-0.964$
according to GCG model.

\end{abstract}


\maketitle

\section{$\text{Introduction}$}

 {\small {~~~}}~The type Ia supernovae (SNe Ia)
investigations \cite{SNe}, the cosmic microwave background(CMB)
results from WMAP \cite{CMB} observations, and surveys of galaxies
\cite{LSS} all suggest that the expansion of present universe is
speeding up rather than slowing down. The accelerated expansion of
the present universe is usually attributed to the fact that dark
energy (DE) is an exotic component with negative pressure. Many
kinds of DE models have already been constructed such as
$\Lambda$CDM \cite{LCDM}, quintessence \cite{quintessence}, phantom
\cite{phantom}, quintom \cite{quintom}, generalized Chaplygin gas
(GCG) \cite{GCG}, modified Chaplygin gas \cite{MCG}, holographic
dark energy \cite{holographic}, agegraphic dark
energy\cite{agegraphic}, and so forth. In addition,
model-independent method\footnote{For example, using mathematical
fundament  one expands  equation of state  of DE $w_{de}$ or
deceleration parameter $q$ with respect to scale factor $a$ or
redshit $z$, such as  $w_{de}(z)=w_{0}$=const \cite{independent1},
$w_{de}(z)=w_{0}+w_{1}z$\cite {independent2},
$w_{de}(z)=w_{0}+w_{1}\ln(1+z)$ \cite{independent3},
$w_{de}(z)=w_{0}+\frac{w_{1}z}{1+z}$ \cite{independent4},
$q(z)=q_{0}+q_{1}z$ \cite{independent1}, $q(z)=q_{0}+
\frac{q_{1}z}{1+z}$ \cite{independent5}, and so forth. Where
$w_{0}$, $w_{1}$, or $q_{0}$, $q_{1}$ are model parameters. For more
information about model-independent method, please see review paper
\cite{review}. } and modified gravity theories (such as
scalar-tensor cosmology \cite{Scalar}, braneworld models
\cite{braneworld})  to interpret accelerating universe have also
been discussed. So, a general and model-independent manner to
distinguish these models introduced by different theories or methods
is necessary. Statefinder diagnostic method is presented in Refs.
\cite{rs}, and it
 has been applied to a large number of DE models
\cite{rspapers}. Recently, another geometrical diagnostic $Om$ is
also introduced in Ref. \cite{Om} to differentiate $\Lambda$CDM with
other models.
 An important property for $Om$ diagnostic
is that it  can be used to distinguish DE models with
 small influence from density parameter $\Omega_{0m}$, though the current observations suggest an uncertainties of at least 25\%
in the value of current matter density $\Omega_{0m}$
\cite{5yWMAPSNeBAO}.  In this paper, we apply $Om$  diagnostic to
GCG model.

The paper is organized as follows. In section 2, the GCG model  as
the
 unification of dark matter and dark energy is introduced briefly. Based on the recently
observed data: the Union SNe Ia \cite{307Union}, the observational
Hubble data (OHD) \cite{OHD}, the baryon acoustic oscillation (BAO)
peak from Sloan Digital Sky Survey (SDSS) \cite{SDSS} and the
five-year WMAP CMB shift parameter \cite{5yWMAP},  $Om$
 diagnostic  is used to GCG model  in section 3. Section
4 is the conclusions.

\section{$\text{generalized Chaplygin gas model}$} \label{section2}

 In the GCG approach, the most interesting property is that the
unknown dark sections in universe--dark energy and dark matter, can
be unified by using an exotic equation of state. The energy density
$\rho$ and pressure $p$ are related by the equation of state (EOS)
\cite{GCG}
\begin{equation}
p=-\frac{A}{\rho^{\alpha}},\label{1}
\end{equation}
where $A$  and $\alpha$ are parameters in the model.

By using the energy conservation equation: $d(\rho
a^{3})=-pd(a^{3})$, the energy density of GCG is expressed as
\begin{equation}
\rho _{GCG}=\rho _{0GCG}[A_s+(1-A_s)(1+z)^{3(1+\alpha )}]^{\frac
1{1+\alpha }},\label{2}
\end{equation}
where $a$ is the scale factor, A$_s=\frac A{\rho _0^{1+\alpha }}$.
For the GCG model, as a scenario of the unification of  dark matter
and dark energy, the GCG fluid is decomposed into two components:
the dark energy component and the dark matter component, i.e., $
\rho _{GCG}=\rho _{de}+\rho _{dm}$, $p_{GCG}=p_{de}$. Then according
to the general recognition about  dark matter, $\rho _{dm}=\rho
_{0dm}(1+z)^3$, the energy density of the DE in the GCG model is
given by
\begin{equation}
\rho _{de}=\rho_{GCG} -\rho _{dm}=\rho
_{0GCG}[A_s+(1-A_s)(1+z)^{3(1+\alpha )}]^{\frac 1{1+\alpha }}-\rho
_{0dm}(1+z)^3.\label{4}
\end{equation}
Furthermore, considering spatially flat  FRW
(Friedmann-Robertson-Walker)
  universe  with  baryon
matter $\rho _{b}$ and GCG fluid $\rho _{GCG}$,  the equation of
state  of DE can be derived as
\begin{equation}
w _{de}=\frac{p_{de}}{\rho_{de}}=\frac{-(1-\Omega
_{0b})A_s[A_s+(1-A_s)(1+z)^{3(1+\alpha )}]^{-\frac \alpha {1+\alpha
}}}{ (1-\Omega _{0b})[A_s+(1-A_s)(1+z)^{3(1+\alpha )}]^{\frac
1{1+\alpha }}-\Omega _{0dm}(1+z)^3},\label{5}
\end{equation}
where $\Omega_{0dm}$ and $\Omega_{0b}$ are present values of
dimensionless dark matter density and baryon matter component. And
 Hubble parameter $H$ is
\begin{equation}
H^2=\frac{8\pi G \rho_{t}}{3}=H_0^2E^{2}=H_0^2\{(1-\Omega
_{0b})[A_s+(1-A_s)(1+z)^{3(1+\alpha )}]^{\frac 1{1+\alpha }}+\Omega
_{0b}(1+z)^3\}.\label{6}
\end{equation}
  $H_{0}$ denotes the current value of Hubble parameter.

\section{$\text{ $Om$ diagnostic for GCG model}$} \label{section3}

It is well known that model-independent quantity $H(z)$ is very
important for understanding the properties of DE, since its value
can be directly obtained from cosmic observations ( for example, the
relation between luminosity distance $D_{L}$ and Hubble parameter is
$H(z)=[\frac{d}{dz}({\frac{D_{L}(z)}{1+z}})]^{-1}$ \cite{H1}
 for
SNe investigations). Recently, a new diagnostic of dark energy $Om$
is introduced to differentiate $\Lambda$CDM with other  dynamical
models. The starting point for $Om$ diagnostic is  Hubble parameter,
and it is defined as \cite{Om}
\begin{equation}
Om(z)\equiv\frac{E^{2}(z)-1}{x^{3}-1},~~x=1+z.\label{23Om}
\end{equation}
Since $Om(z)$ only depends upon the scale factor $a$ and its
derivative, it is a "geometrical" diagnostic. For $\Lambda$CDM
model, $Om(z)=\Omega_{0m}$ is a constant, then it provides a null
test of this model\footnote{For null test of $\Lambda$CDM model, one
can also see Ref.  \cite{testLCDM}.}. The benefit of $Om$ diagnostic
is that the quantity $Om(z)$ can distinguish DE models with less
dependence on matter density $\Omega_{0m}$ relative to the EOS of DE
$w_{de}(z)$ \cite{Om}.

In what follows, we use a combination of the recent standard candle
data (Union SNe Ia \cite{307Union}) and the OHD  to constrain  the
evolutions of $Om(z)$ and $w_{de}(z)$ for GCG model.
  The Union SNe  data
includes the SNe samples from the Supernova Legacy Survey
\cite{SNLS}, ESSENCE Surveys \cite{ESSENCE}, distant SNe discovered
by the Hubble Space Telescope \cite{HST}, nearby SNe \cite{nearby}
 and several other, small
data sets \cite{307Union}. The OHD  are given by calculating the
differential ages of passively evolving galaxies from the GDDS
\cite{OHD1} and archival data \cite{OHD2}. According to the
expression $H(z)=-\frac{1}{1+z}\frac{dz}{dt}$, one can see that the
value of $H(z)$ can be directly obtained by the determination of the
differential age $dz/dt$.  Ref. \cite{OHD} get nine values of $H(z)$
in the range of $0<z<1.8$ (see Table 1). And these nine
observational Hubble data have been used to constrain  DE
 models
 \cite{OHDpaper}.

\begin{table}
\vspace*{-12pt}
\begin{center}
\begin{tabular}{c | c  }
\hline\hline $z$
&~0.09~~~~~~0.17~~~~~~0.27~~~~~~0.40~~~~~~0.88~~~~~~1.30~~~~~~1.43~~~~~~1.53~~~~~~1.75~~~~~~
\\\hline
   $H(z)$ (kms$^{-1}$~Mpc)$^{-1}$   &~~~69~~~~~~~~83~~~~~~~70~~~~~~~~~~87~~~~~~~117~~~~~~~168~~~~~~~177~~~~~~~140~~~~~~202~~~~~~
 \\\hline
    $1\sigma$ uncertainty
    &~~$\pm$12~~~~~$\pm$8.3~~~~~~$\pm$14~~~~$\pm$17.4~~~~$\pm$23.4~~~~$\pm$13.4~~~~$\pm$14.2~~~~~$\pm$14~~~~$\pm$40.4~~~~~~
  \\\hline\hline
       \end{tabular}
       \end{center}
        Table 1. The observational $H(z)$ data
        \cite{OHD3}\cite{OHDpaper}.
       \end{table}

\begin{figure}[!htbp]
\includegraphics[width=4cm]{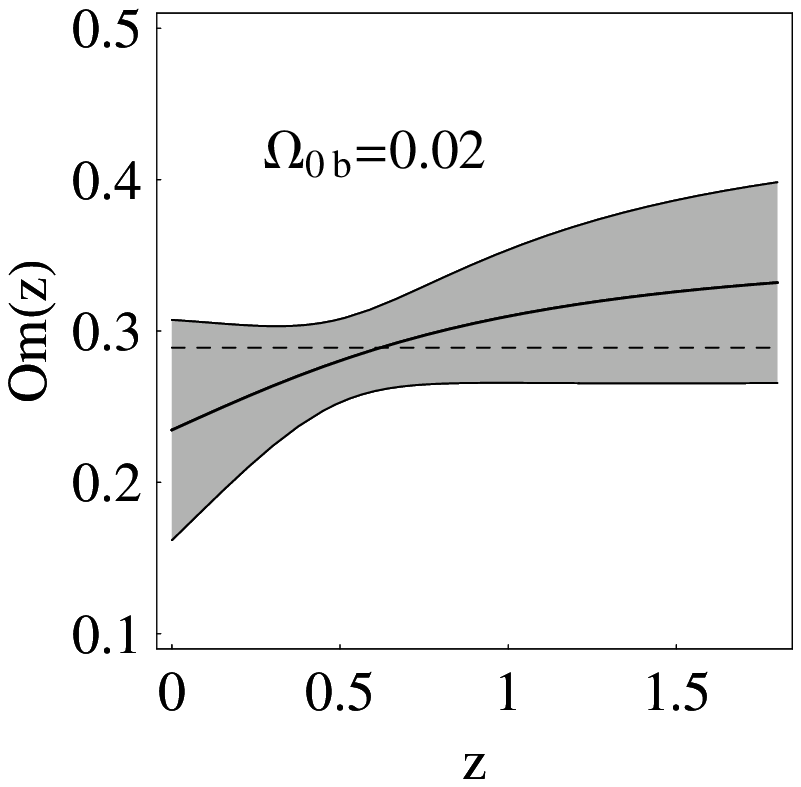}~~~
\includegraphics[width=4cm]{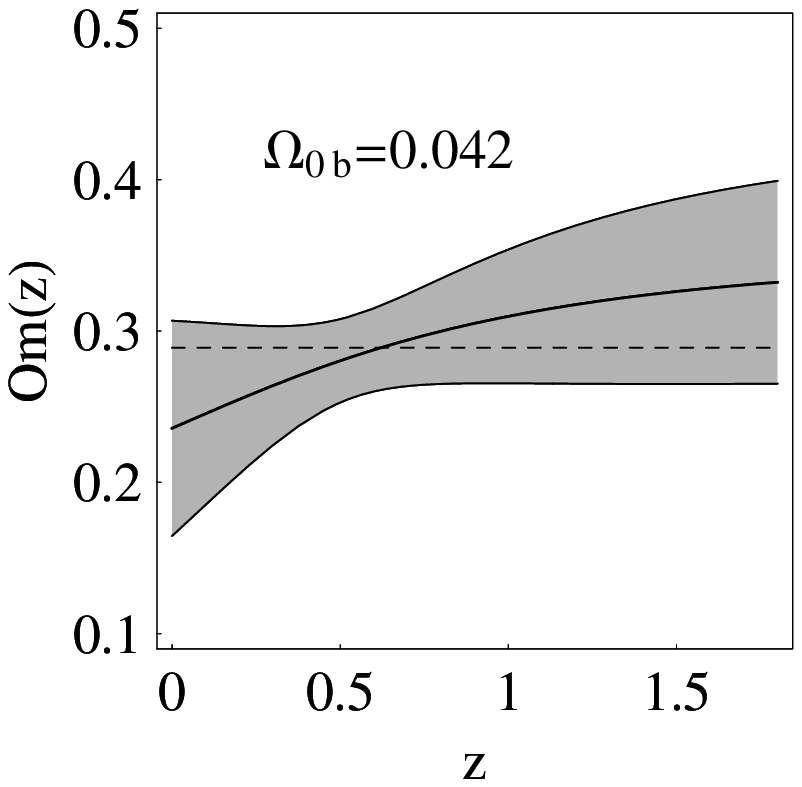}~~~
~\includegraphics[width=4cm]{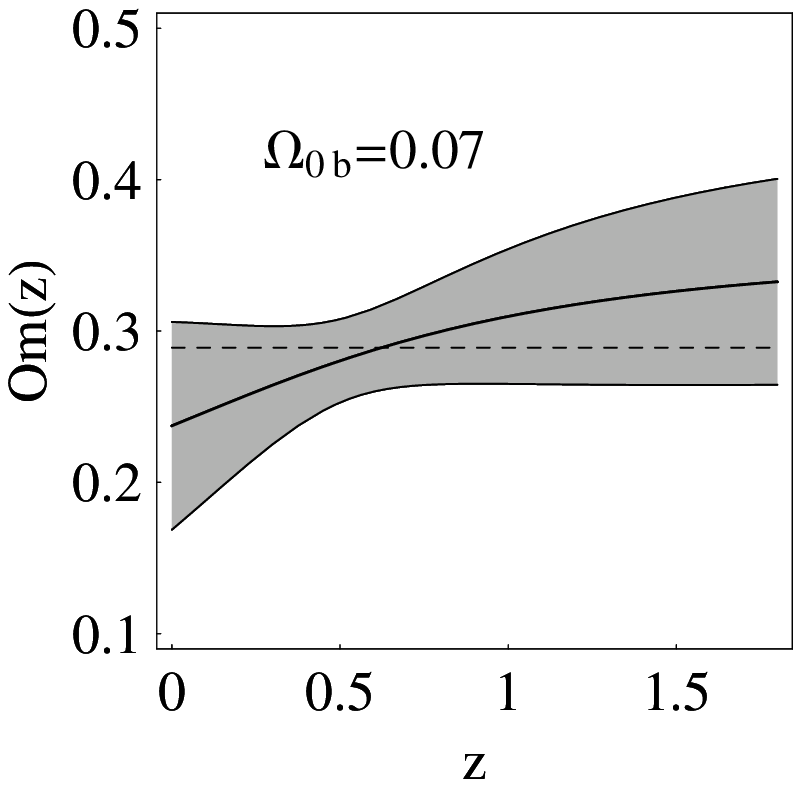}\\
\includegraphics[width=4cm]{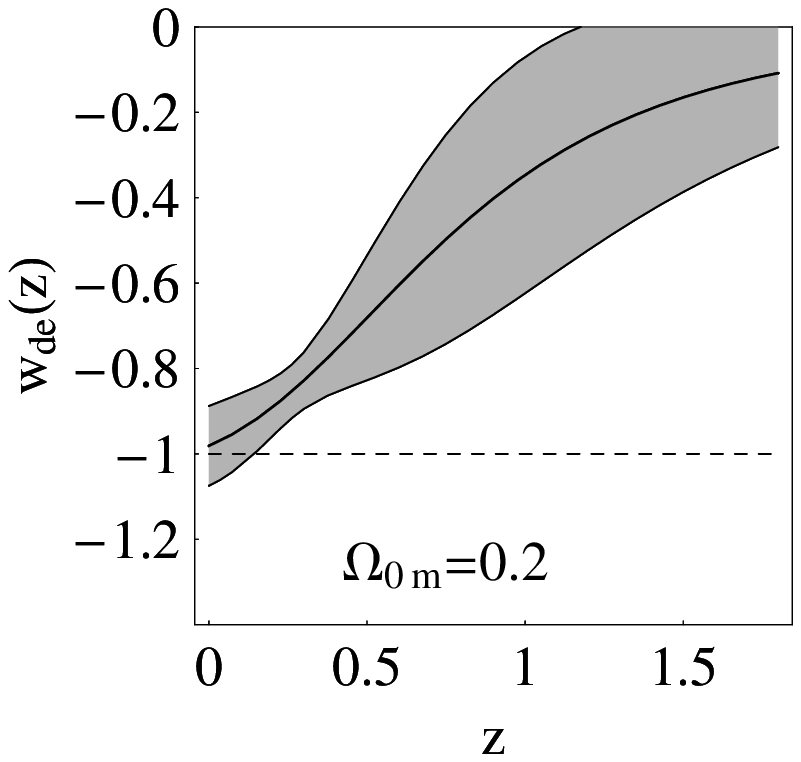}~~~
\includegraphics[width=4cm]{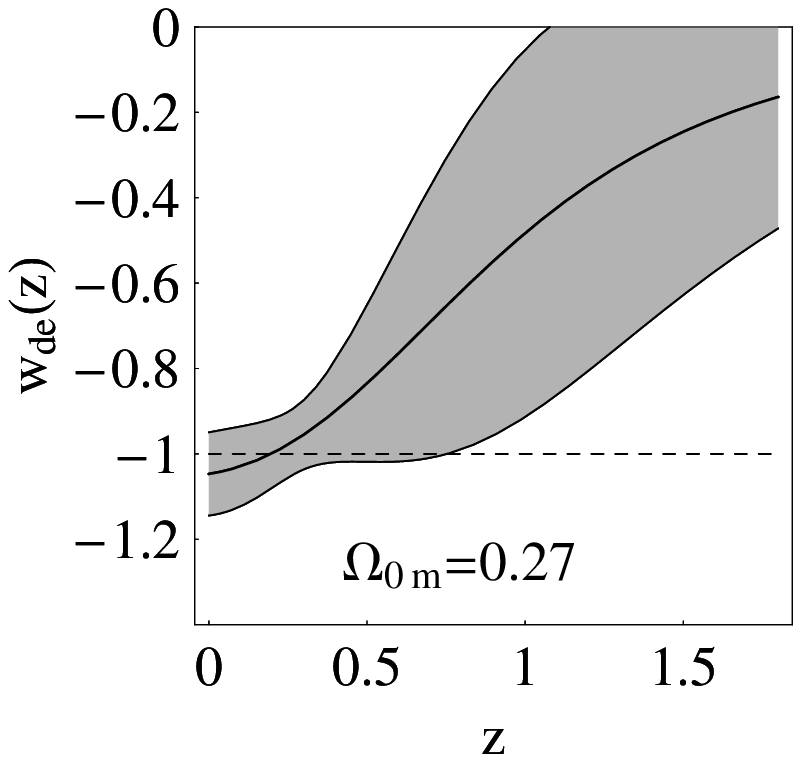}~~
~\includegraphics[width=4cm]{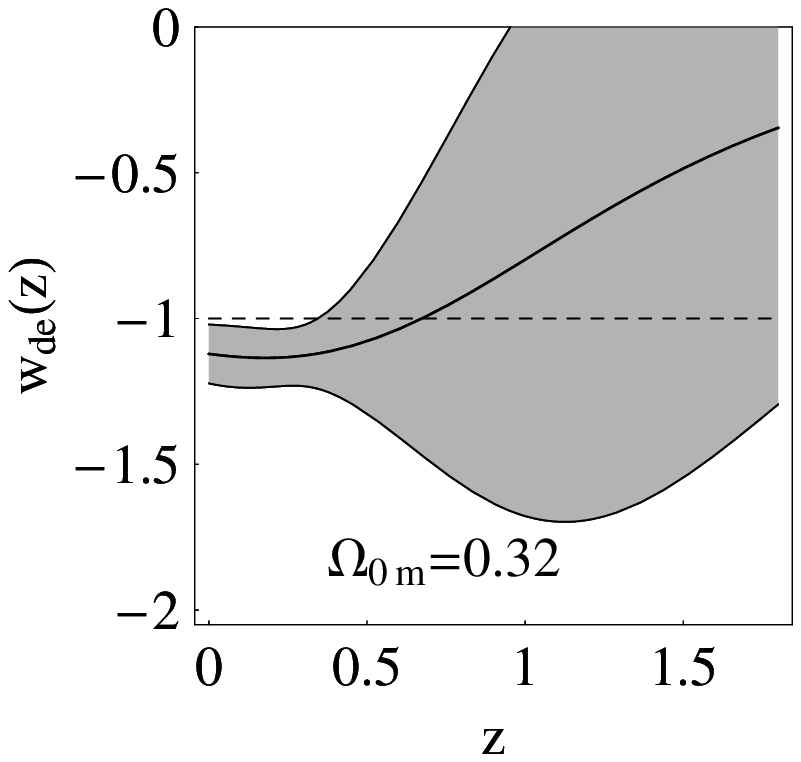}
 \caption{ Evolutions of $Om(z)$ and $w_{de}(z)$ by using a combination of Union SNe data and OHD  for GCG
model. Here three different values $\Omega_{0b}$=0.02, 0.042, 0.07
 for $Om(z)$ evolution diagram, and
$\Omega_{0m}=\Omega_{0b}+\Omega_{0dm}$=0.22, 0.27, 0.32 for
$w_{de}(z)$ diagram are assumed. The shaded regions show the
1$\sigma$ confidence level. The dashed lines show the values of
$Om(z)$ and $w_{de}(z)$ for $\Lambda$CDM model.}\label{Omwdefigure}
\end{figure}

 From Eq. (\ref{5}),  it can be seen that both
$\Omega_{0b}$ and $\Omega_{0dm}$ are included in the expression of
 $w_{de}(z)$  for GCG model. Given three different
values of $\Omega_{0m}$, the evolutions of $w_{de}(z)$ with
$1\sigma$ confidence level for GCG model are plotted in Fig.
\ref{Omwdefigure} (lower) by using the Union SNe data and the OHD.
 Furthermore according to Eq. (\ref{6}), we can see that the Hubble
parameter $H(z)$ for GCG model is dependent on the baryon density
$\Omega_{0b}$ and two model parameters ($A_{s},\alpha$). It does not
explicitly include current matter density $\Omega_{0m}$. And one
knows that the observational constraints on  parameter $\Omega_{0b}$
is more stringent\footnote{Such
 as $\Omega_{0b}h^{2}=0.0214\pm 0.0020$ from the observation of the deuterium to hydrogen ratio
 towards QSO absorption systems \cite{QSOob0},
and $\Omega_{0b}h^{2}=0.02273\pm0.00062$ from the five-year WMPA
 results for the observation of CMB \cite{5yWMAP},
here $h=H_{0}/100$.},
 i.e., it has a relative smaller variable range  relative to $\Omega_{0m}$.
On the basis of Eq. (\ref{23Om}), we plot the evolutions of $Om(z)$
for GCG model in Fig. \ref{Omwdefigure} (upper). From Fig.
\ref{Omwdefigure}, it can be found
  that the
$Om(z)$ diagram for GCG model as the unification of dark matter and
dark energy is almost independent of the variation of $\Omega_{0b}$,
but the evolution of $w_{de}(z)$ is sensitive to the variation  of
matter density.

In Ref. \cite{Ompaper}, $Om$ diagnostic has been used to distinguish
$\Lambda$CDM and Ricci DE model.  Assuming the matter density
$\Omega_{0m}$ to be a free parameter,  based on the recent cosmic
observations  Ref. \cite{Om} plots the evolution diagram of $Om(z)$
in a model-independent CPL scenario\footnote{It is an expansion for
EOS of DE relative to scale factor $a$,
$w_{de}(a)=w_{0}+w_{1}(1-a)$, or
$w_{de}(z)=w_{0}+\frac{w_{1}z}{1+z}$
\cite{independent4}\cite{indep4}.}. In this paper, treating
$\Omega_{0b}$ as a free parameter, we apply the $Om$ diagnostic to
GCG model. One knows that for the same dark energy model, the
different evolutions of cosmological quantity can be obtained from
different observational datasets. This is the so-called
data-dependent. And in order to diminish systematic uncertainties
and get the stringent constraint on cosmological parameters, people
often combine many observations to constrain the evolutions of
cosmological quantities.
  Next we use a combination of the  recent standard candle data,  the standard ruler data
(the BAO peak from SDSS and the five-year WMAP CMB shift parameter
$R$) and the OHD  to constrain the evolution of $Om(z)$ for GCG
model.

Because the universe has a fraction of baryons, the acoustic
oscillations in the relativistic plasma would be imprinted onto the
late-time power spectrum of the non-relativistic matter
\cite{27Eisenstein}.
 Then the observations of acoustic signatures in
the large-scale clustering of galaxies can be used to constrain  DE
models with detection of a peak. The measured data at $z_{BAO}=0.35$
from SDSS is \cite{SDSS}
\begin{equation}
A=\sqrt{\Omega_{0m}^{eff}}E(z_{BAO})^{-1/3}[\frac{1}{z_{BAO}}\int_{0}^{z}\frac{dz^{'}}{E(z^{'})}]^{2/3}=0.469\pm0.017,\label{9A-BAO}
\end{equation}
where $\Omega_{0m}^{eff}$ is the effective matter density parameter
\cite{GCG1}.

The structure of the anisotropies of the cosmic microwave background
radiation depends on two eras in cosmology, i.e., the last
scattering era and today. They can also be applied to limit  DE
models by using the shift parameter \cite{RCMB}
\begin{equation}
R=\sqrt{\Omega_{0m}^{eff}}\int_{0}^{z_{rec}}\frac{H_{0}dz^{'}}{H(z^{'})}=1.715\pm0.021,\label{6R-CMB}
\end{equation}
where $z_{rec}=1089$ is the redshift of recombination, and the value
of $R$ is given by five-year WMAP data \cite{5yWMAP}\cite{Om}.

\begin{figure}[!htbp]
\includegraphics[width=5cm]{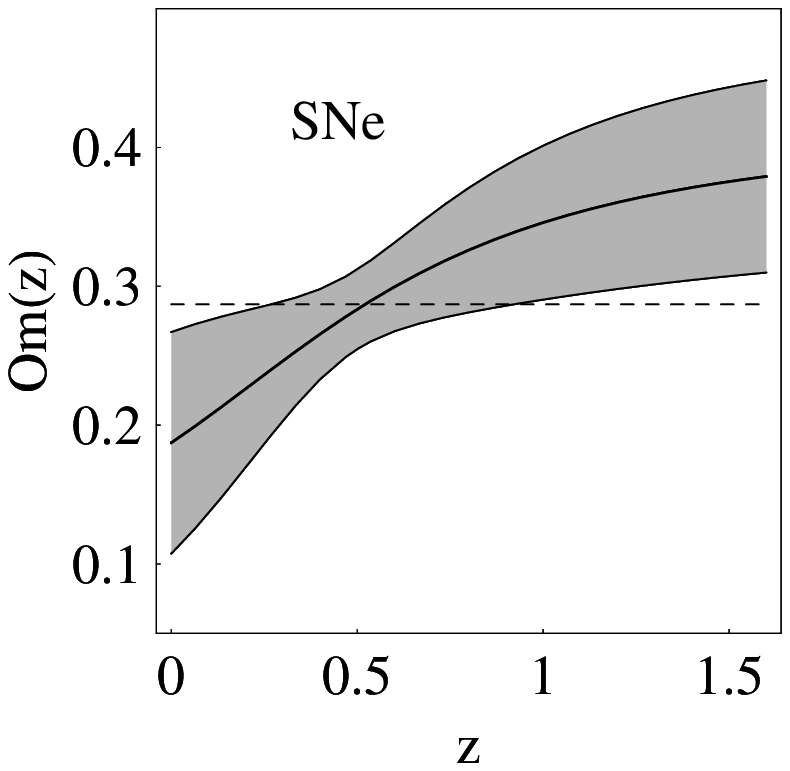}~~~~~~~~~~~~~~~
\includegraphics[width=5cm]{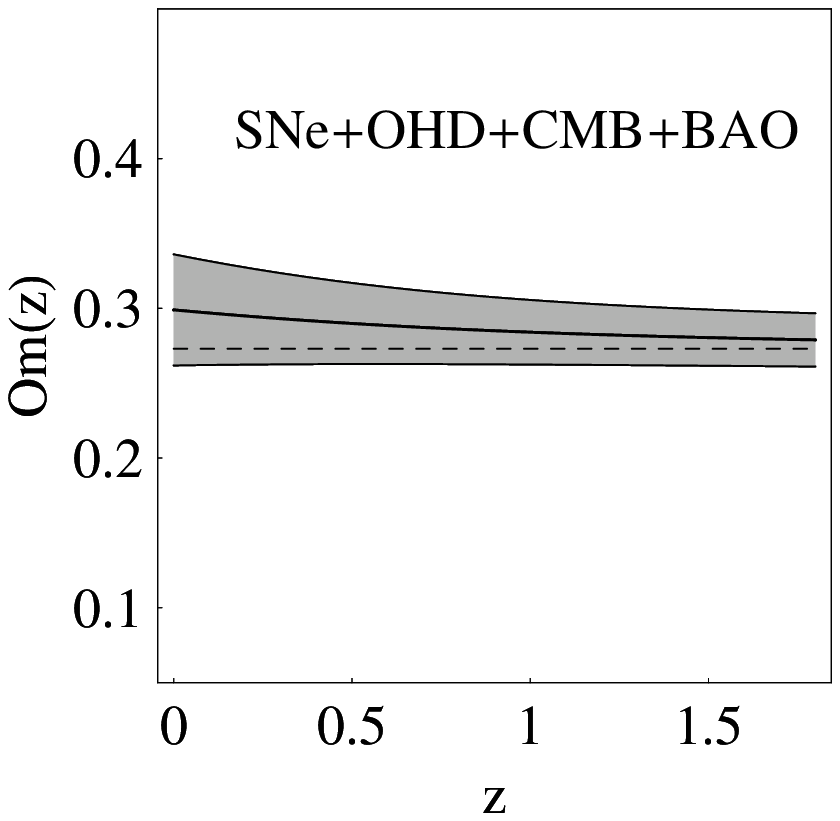}\\
~~~~~~~~~~~(a)~~~~~~~~~~~~~~~~~~~~~~~~~~~~~~~~~~~~~~~~~~~~~~~~~~~~~~~~~~~(b)~~\\
 \caption{ $Om(z)$ diagnostic for GCG model from  Union SNe data  and a
combination of Union SNe, OHD, CMB and BAO data.  Here $\Omega_{0b}$
is treated as a free parameter. The shaded regions show the
1$\sigma$ confidence level. The dashed lines show the values of
$Om(z)$ for $\Lambda$CDM model obtained from the corresponding
observational constraint.}\label{Omfigure}
\end{figure}

 We plot the
evolution of $Om(z)$ for GCG model by using the single standard
candle data in Fig. \ref{Omfigure} (a).  From this figure, it is
easy to see that the difference between GCG and $\Lambda$CDM model
is obvious. Since the $Om$ diagnostic is relatively insensitive to
the  density parameter, the difference between this figure and  Fig.
1 (upper) is caused by using the different datasets to constrain the
quantity $Om(z)$, i.e. figure 1 is determined from a combination of
Union SNe Ia and OHD data, but figure 2 (a) is plotted by means of
Union SNe Ia data alone. Furthermore based on above four
observational datasets, the combined constraint on $Om(z)$  is
presented in Fig. \ref{Omfigure} (b). According to
 Fig. \ref{Omfigure} (b), we can see that the best fit
evolution of $Om(z)$ for GCG model is  near to $\Lambda$CDM case,
and $Om(0)\equiv Om(z=0)=0.299~^{+0.037}_{-0.037}~(1\sigma)$ for GCG
model. In addition, by using above four  datasets to $\Lambda$CDM
model, it is obtained that the best fit value of $\Omega_{0m}$ with
confidence level is
$\Omega_{0m}=0.273~^{+0.016}_{-0.015}~(1\sigma)$. We know
$Om(z)=\Omega_{0m}$ for $\Lambda$CDM, then  its best fit evolution
is included in the $1\sigma$ confidence level of $Om(z)$
 in GCG scenario. And it can be seen that  at 1$\sigma$
confidence level, these two models can not be clearly distinguished
by current observed data according to the $Om(z)$ diagram.

 At last, according to the expression
$\frac{Om(z)-\Omega_{0m}}{1-\Omega_{0m}}\simeq 1+w_{0de}$ ($z\ll1$)
\cite{Om}, it can be calculated that the current EOS of DE
$w_{0de}\simeq -0.964$ by taking $\Omega_{0m}=0.273$ and the best
fit value $Om(0)=0.299$.

\section{$\text{Conclusion}$} \label{section4}

 On the basis
of the recently observed data: the Union SNe Ia data, the nine
observational Hubble data, the SDSS baryon acoustic peak and the
five-year WMAP result, we apply a geometrical diagnostic $Om$ to
distinguish GCG model and $\Lambda$CDM model. From Fig.
\ref{Omwdefigure}, it is shown that the larger error for  the
evolution of $w_{de}(z)$ may be produced by the erroneous estimation
of matter density $\Omega_{0m}$. And the $Om(z)$ is a better
quantity than $w_{de}(z)$ to truly distinguish DE models and to show
the properties of DE. According to the $Om$ diagram, it is easy to
see that for the constraint from the single standard candle data,
the difference between GCG model and $\Lambda$CDM model is obvious,
while for the combined constraint, the best fit evolutions of
$Om(z)$ for them are similar and the difference between these two
models is not clear at $1\sigma$ confidence level. In addition, we
also calculate the value of current EOS of DE, $w_{0de}=-0.964$, by
using the value of $Om(0)$ for GCG model. Here the $Om(z)$ diagram
is not sensitive to the variation of  density parameter.

\textbf{\ Acknowledgments } The research work is supported by NSF
(10703001) of PR China.


\begin{thebibliography}{*}
\bibitem{SNe} A.G. Riess {\it et al}, 1998 {\it Astron. J.} {\bf 116} 1009
[arXiv:astro-ph/9805201]\\
S. Perlmutter {\it et al}, 1999 {\it Astrophys. J.} {\bf  517} 565



\bibitem{CMB} D.N. Spergel {\it et al},  2003 {\it Astrophys. J. Suppl.} {\bf 148} 175
[arXiv:astro-ph/0302209]


\bibitem{LSS} A.C. Pope {\it et al},  2004 {\it Astrophys. J.}  {\bf 607} 655
[arXiv:astro-ph/0401249]

\bibitem{LCDM} S. Weinberg, 1989 {\it Mod. Phys. Rev.} {\bf 61} 527


\bibitem{quintessence} B. Ratra and P.J.E. Peebels, 1988 {\it Phys. Rev. D.} {\bf 37} 3406

\bibitem{phantom}  R.R. Caldwell, M. Kamionkowski and N. N.
Weinberg, 2003 {\it Phys. Rev. Lett.} {\bf91} 071301
[arXiv:astro-ph/0302506]\\
M.R. Setare, 2007 {\it Eur. Phys. J. C }{\bf 50} 991


\bibitem{quintom} B. Feng, X.L. Wang and X.M. Zhang, 2005  {\it Phys. Lett. B} {\bf 607} 35
[arXiv:astro-ph/0404224]


\bibitem{GCG} A.Y. Kamenshchik, U. Moschella and V. Pasquier, 2001 {\it Phys. Lett. B} {\bf 511} 265
[arXiv:gr-qc/0103004]\\
M.C. Bento, O. Bertolami and A.A. Sen, Phys. Rev. D 66 (2002) 043507 [arXiv:gr-qc/0202064]\\
 P.X. Wu and
H.W. Yu  2007 {\it Phys. Lett. B} {\bf 644} 16


\bibitem{MCG} H.B. Benaoum, [arXiv:hep-th/0205140]\\
 S. Li, Y.G. Ma and Y. Chen, [arXiv:astro-ph/0809.0617]\\
J.B. Lu {\it et al},  2008 {\it Phys. Lett. B}  {\bf 662}, 87





\bibitem{holographic} M. Li, 2004 {\it Phys. Lett. B}  {\bf 603} 1
[arXiv:hep-th/0403127]\\
Q. Wu, Y.G. Gong, A.Z. Wang and J.S. Alcanizd, 2008  {\it Phys.
Lett. B}  {\bf 659} 34\\
M R Setare, 2007 {\it Phys.Lett.B}  {\bf 648} 329
 [arXiv:hep-th/0704.3679]



\bibitem{agegraphic} R.G. Cai, 2007 {\it Phys. Lett. B} {\bf 657} 228  [arXiv:hep-th/0707.4049]



 \bibitem{independent1} A.G. Riess  {\it et al}, 2004 {\it Astrophys. J.} {\bf 607} 665  [arXiv:astro-ph/0402512]


\bibitem{independent2}  A.R. Cooray and D. Huterer, 1999 {\it Astrophys. J.} {\bf 513}  L95
[arXiv:astro-ph/9901097]\\
J.V. Cunha, L. Marassi and R.C. Santos, 2007  {\it Int. J. Mod.
Phys. D} {\bf 16} 403


\bibitem{independent3}  B.F. Gerke and G. Efstathiou,  2002 {\it Mon. Not. R. Astron. Soc.} {\bf 335} 33
 [arXiv:astro-ph/0201336]

\bibitem{independent4}    E.V. Linder, 2003 {\it Phys. Rev. Lett.} {\bf 90}  091301
astro-ph/0208512\\
M. Chevallier and D. Polarski,  2001 {\it Int. J. Mod. Phys. D} {\bf
10} 213 [arXiv:gr-qc/0009008]


\bibitem{independent5} L.X. Xu and J.B. Lu, 2009 {\it Mod. Phys. Lett. A}  {\bf 24}
369

\bibitem{review} V. Sahni and A. A. Starobinsky, 2006 {\it Int. J. Mod. Phys. D} {\bf 15} 2105
[arXiv:astro-ph/0610026]

\bibitem{Scalar} B. Boisseau, G. Esposito-Farese, D. Polarski and A. A.
Starobinsky, 2000 {\it Phys. Rev. Lett.} {\bf 85} 2236
[arXiv:gr-qc/0001066]\\
M. Trodden,  2007 {\it Int. J. Mod. Phys. D} {\bf 16} 2065

\bibitem{braneworld}
G. Dvali, G. Gabadadze  and M. Porrati, 2000  {\it Phys. Lett. B}
{\bf 485}  208
 [arXiv:hep-th/0005016]\\
  V. Sahni and Yu. Shtanov, 2003 {\it J. Cosmol. Astropart. Phys.} {\bf 0311}  014
  [arXiv:astro-ph/0202346]\\
   V. Sahni, Yu. Shtanov and A. Viznyuk, 2005
{\it J. Cosmol. Astropart. Phys.} {\bf 0512} 005
[arXiv:astro-ph/0505004]\\
I. Brevik, 2008 {\it Eur. Phys. J. C } {\bf 56} 579


\bibitem{rs} V. Sahni and T.D. Saini, A. A. Starobinsky and U. Alam, 2003 {\it JETP Lett.} {\bf
77}  201-206

\bibitem{Om} V. Sahni, A. Shafieloo and A. A. Starobinsky, 2008 {\it  Phys. Rev. D} {\bf 78} 103502
[arXiv:astro-ph/0807.3548]


\bibitem{rspapers} U. Alam, V. Sahni, T. D. Saini and A. A.
Starobinsky [arXiv:astro-ph/0303009]\\
W. Zhao, 2008 {\it Int. J. Mod. Phys. D} {\bf 17} 1245\\
 X. Zhang, F.Q. Wu and J.F. Zhang,  2006 {\it J. Cosmol. Astropart. Phys.}  {\bf0601} 003   [arXiv:astro-ph/0411221]


\bibitem{5yWMAPSNeBAO} E. Komatsu {\it et al},
[arXiv:astro-ph/0803.0547]


\bibitem{307Union} D. Rubin  {\it et al},
[arXiv:astro-ph/0807.1108]


\bibitem{OHD} J. Simon   {\it et al},  2005  {\it Phys. Rev. D} {\bf71}, 123001

\bibitem{SDSS} D.J. Eisenstein {\it et al},  2005 {\it Astrophys. J.} {\bf 633}, 560
[arXiv:astro-ph/0501171]


\bibitem{5yWMAP} J. Dunkley {\it et al},  [astro-ph/0803.0586]




\bibitem{H1}
T. Nakamura and T. Chiba, 1999 {\it Mon. Not. R. Astron. Soc.} {\bf
306} 696 [arXiv:astro-ph/9810447]\\
A.A. Starobinsky, 1998 {\it JETP Lett.}  {\bf68} 757
[arXiv:astro-ph/9810431]


\bibitem{testLCDM} C. Zunckel and C. Clarkson, 2008 {\it Phys. Rev. Lett.} {\bf
101} 181301  [arXiv:astro-ph/0807.4304]\\
T. Chiba and T. Nakamura, 2007 {\it Prog. Theor. Phys.} {\bf118} 815
  [arXiv:astro-ph/0708.3877]



\bibitem{SNLS} P. Astier  {\it et al},  2006 {\it Astron. Astrophys.} {\bf447} 31
[arXiv:astro-ph/0510447]


\bibitem{ESSENCE} W.M. Wood-Vasey  {\it et al},  [arXiv:astro-ph/0701041]




\bibitem{HST} A.G. Riess  {\it et al},  [arXiv:astro-ph/0611572]

\bibitem{nearby} M. Hamuy, M.M. Phillips, N.B. Suntzeff, R.A. Schommer and J. Maza,
1996 {\it Astron. J.} {\bf 112} 2408  [arXiv:astro-ph/9609064]\\
S. Jha, A. G. Riess and R. P. Kirshner, 2007 {\it Astrophys. J.}
{\bf 659} 122  [arXiv:astro-ph/0612666]



\bibitem{OHD1} R.G. Abraham  {\it et al},  2003  {\it Astron. J.} {\bf593}
622


\bibitem{OHD2} T. Treu   {\it et al},  1999 {\it Mon. Not. R. Astron. Soc.} {\bf308} 1037
\\
T. Treu  {\it et al},  2001 {\it Mon. Not. R. Astron. Soc.} {\bf326}
221



\bibitem{OHD3} L. Samushia and B. Ratra, 2006 {\it Astrophys. J.} {\bf650} L5  [astro-ph/0607301]\\
 R. Jimenez, L. Verde, T. Treu and D. Stern, 2003 {\it Astrophys. J.} {\bf593} 622
 [astro-ph/0302560]

\bibitem{OHDpaper} Z.L. Yi and T.J. Zhang, 2007 {\it Mod. Phys.
Lett. A} {\bf 22} 41-53 [arXiv:astro-ph/0605596]\\
H. Wei, S. N. Zhang, 2007 {\it Phys. Lett. B}  {\bf644} 7
[astro-ph/0609597]\\
J.B Lu, L.X Xu, M.L Liu and Y.X Gui, 2008 {\it Eur. Phys. J. C}
{\bf58} 311 [arXiv:astro-ph/0812.3209]\\
Hui Lin {\it et al}, [arXiv:astro-ph/0804.3135]

\bibitem{QSOob0}  D. Kirkman {\it et al},  2003 {\it Astrophys. J. Suppl.} {\bf 149}   1
[arXiv:astro-ph/0302006]



\bibitem{Ompaper} C.J. Feng, [arXiv:astro-ph/0809.2502]


\bibitem{indep4} E. M. Barboza Jr. and J. S. Alcaniz, 2008 {\it Phys. Lett. B} {\bf 666}  415-419 [arXiv:astro-ph/0805.1713]


\bibitem{27Eisenstein} D.J. Eisenstein and W. Hu, 1998 {\it Astrophys. J.} {\bf 496} 605
[arXiv:astro-ph/9709112]

\bibitem{GCG1} M. Makler, S.Q. Oliveira and I. Waga, 2003 {\it Phys. Rev. D} {\bf 68}
123521\\
J.A.S. Lima, J.V. Cunha and J.S. Alcaniz,
[arXiv:astro-ph/0611007]\\
Z.H. Zhu, 2004 {\it Astron. Astrophys.} {\bf 423}  421


\bibitem{RCMB} J.R. Bond, G. Efstathiou and M. Tegmark,  1997 {\it Mon. Not. R. Astron. Soc.} {\bf 291}, L33
[arXiv:astro-ph/9702100]














\end{thebibliography}
\end{document}